\documentclass[aip,jcp,amsmath,amssymb,showkeys,floatfix,
reprint
]{revtex4-1}

\usepackage{float}
\usepackage{placeins}
\usepackage{color}
\usepackage[OT1]{fontenc}
\usepackage[utf8]{inputenc}
\usepackage{graphicx}
\usepackage{braket}
\usepackage{siunitx}
\usepackage{amsmath}
\usepackage{amssymb}
\usepackage{bm}
\usepackage{booktabs}
\usepackage[capitalize]{cleveref}
\usepackage{placeins}
\usepackage{droidsans}
\usepackage{soul}
\usepackage[normalem]{ulem} 
\usepackage{appendix}

\begin{document}

\title{Comment on ``Regularizing the MCTDH equations of motion
through an optimal choice on-the-fly (i.e., spawning) of unoccupied
single-particle functions'' [D. Mendive-Tapia, H.-D. Meyer, J. Chem. Phys. 153, 234114 (2020)]}

\author{Rocco Martinazzo}
\email{rocco.martinazzo@unimi.it}
\affiliation{Department of Chemistry, Universit{\`a} degli Studi di Milano, Via Golgi 19, 20133 Milano, Italy}
\author{Irene Burghardt}
\email{burghardt@chemie.uni-frankfurt.de}
\affiliation{Institute for Physical and Theoretical Chemistry, Goethe
University Frankfurt, Max-von-Laue-Str.\ 7, 60438 Frankfurt/Main, Germany}

\date{\today}

\providecommand{\vx}{\vec{R}}
\providecommand{\uM}{\underline{M}^{-1}}
\providecommand{\vp}{\vec{P}}
\providecommand{\nH}{\langle\nabla_{\vec{R}}H\rangle}
\providecommand{\hve}{\hat{\vec{\epsilon}}}

\maketitle

Time-dependent variational methods are a powerful approach to quantum
propagation in many dimensions. This is exemplified by the Multi-Configuration
Time-Dependent Hartree (MCTDH) method \cite{Meyer1990,Beck2000} and its
hierarchical multi-layer variant,\cite{Wang2003,Wang2015} as well as related
approaches like Gaussian-based G-MCTDH\cite{Burghardt1999,Roemer2013} that
connect to a semiclassical trajectory-type picture. These methods rely on
time-evolving basis sets whose equations of motion are determined by a
time-dependent variational principle.\cite{Beck2000,Lubich2005}

Besides the optimal, variational evolution of a specified basis set, an
important problem is the construction of an {\em adaptive} basis where
functions are added or removed depending on the system's time-evolving
correlations. This problem has been typically addressed in an {\em ad hoc}
fashion, especially in the context of Gaussian wavepacket methods where the
notion of spawning during nonadiabatic events was introduced.\cite{Yang2009}
In the context of MCTDH, addition of unoccupied basis functions, notably at
the start of the propagation, requires regularization due to singularities in
the equations of motion. To this end, approaches have been proposed which rely
on short-time perturbative expansions\cite{Manthe2015} or natural orbital
population thresholds.\cite{MendiveTapia2017}

In a recent paper,\cite{MendiveTapia2020} Mendive-Tapia and Meyer present a
modified MCTDH propagation scheme where the basis set is expanded {\em
on-the-fly}, according to a variational error criterion that is augmented by a
contribution of additional, unoccupied basis functions. In the notation of
Ref.\ [\onlinecite{MendiveTapia2020}], the augmented error criterion reads
\begin{eqnarray}  
\tilde{E} & := &  \| \dot{\Psi}_e - \tilde{\dot{\Psi}}_M \|^2 = \min
\label{eq:error-aug-MM}
\end{eqnarray}
where $\dot{\Psi}_e$ fulfills the Schr\"odinger equation and
$\tilde{\dot{\Psi}}_M$ is the time derivative of the augmented MCTDH
wavefunction; in the latter, a set of unoccupied basis functions appear {\em
via} the time derivative of their coefficients and an augmented projector.
Minimizing the augmented error of Eq.\ (\ref{eq:error-aug-MM}) determines the
choice of the optimal unoccupied basis functions in a fully variational
setting. The authors detail the corresponding algorithm and report superior
performance of the new scheme as compared with other spawning approaches, for
a model system as well as a realistic system in six dimensions.

The purpose of the present Comment is to point out that the approach of
Ref.\ [\onlinecite{MendiveTapia2020}] is an instance of a general concept
established in previous work\cite{Martinazzo2020,Martinazzo2019} where
adaptive variational quantum propagation based on the
Local-in-Time Error (LITE) was introduced and exemplified for variational
Gaussian wavepacket dynamics.

The LITE, $\varepsilon_{\cal M}$, is defined as the instantaneous deviation from the exact
solution at the variational minimum,
\begin{eqnarray}
\varepsilon_{\cal M} [\Psi_0] & = & \frac{1}{\hbar} \min\limits_{\dot{\Psi}_0\in
{T}_{0}\mathcal{M}} \| i \hbar \dot{\Psi}_0 - H \Psi_0 \|
\label{eq:LITE1}
\end{eqnarray}
where the time derivative of the variational solution, $\dot{\Psi}_0$, is an
element of the tangent space ${T}_{0}\mathcal{M}$ of the variational manifold
$\mathcal {M}$.\cite{Broeckhove88,Lubich2005,Hackl2020} The LITE refers to the
variationally optimal solution to the short-time dynamics taking the system
from an initial state $\Psi_0$ to a state $\Psi_0(dt)$ (which generally
deviates from the exact solution $\Psi_0^{\rm exact}(dt)$). When resizing the
variational manifold {\em on-the-fly} by spawning, $\dot{\Psi}_0$ in
Eq.\ (\ref{eq:LITE1}) is replaced by $\dot{\Psi}_s$, and the modified LITE
{$\varepsilon_{{\cal M}_s} [\Psi_0]$} is used to optimally determine the
additional basis functions\cite{Martinazzo2020,Martinazzo2019} {(${\cal M}_s$
denotes the extended manifold)}. This is entirely parallel to the arguments
given in Ref.\ [\onlinecite{MendiveTapia2020}], and indeed the augmented error
of Eq.\ (\ref{eq:error-aug-MM}) can be identified as
{$\tilde{E} = \varepsilon^2_{{\cal M}_s}[\Psi_0]$}.

The LITE as defined in Eq.\ (\ref{eq:LITE1}) lies at the heart 
of the McLachlan variational principle (VP)
\cite{McLachlan1964} but it equally applies to the rather common situation
where both $\dot{\Psi}_0$ and $i\dot{\Psi}_0$ lie in ${T}_{0}\mathcal{M}$,
entailing that the Dirac-Frenkel VP can be employed and all the known VPs
become equivalent to each other.\cite{Broeckhove88,Hackl2020} The McLachlan VP
underscores the geometric idea of optimization by a minimum distance criterion
and leads to simplified expressions for the squared LITE,
\begin{eqnarray}
\varepsilon_{\cal M}^2 [\Psi_0] & = & \frac{1}{\hbar^2} ( \|H \Psi_0 \|^2 -
\hbar^2 \| \dot{\Psi}_0 \|^2)
\label{eq:LITE2}
\end{eqnarray}
and for the error reduction due to spawning, $\Delta \varepsilon_s$,
which takes the form
\begin{eqnarray}
\Delta \varepsilon_s^2 & = & \| \dot{\Psi}_s  \|^2 - \| \dot{\Psi}_0 \|^2=\frac{1}{\hbar}
\Im{\braket{\delta_s \dot{\Psi}|H|\Psi_0}}
\label{eq:LITEerror}
\end{eqnarray}
 Here, $\delta_s\dot{\Psi}:=\dot{\Psi}_s-\dot{\Psi}_0$  is chosen such as to
maximize $\Delta \varepsilon_s^2$ to achieve optimal
spawning. The application of this {\em optimal spawning criterion} was demonstrated in
Ref.\ [\onlinecite{Martinazzo2020}] for variational Gaussian wavepacket
dynamics, and an extension to MCTDH was suggested. The connection to the
theoretical developments of Ref.\ [\onlinecite{MendiveTapia2020}] is further detailed
in App.\ \ref{sec:app1}.

The LITE has various important implications for the variational dynamics. It
determines the {\em a posteriori} error bound\cite{Lubich2005} for the
deviation from the exact solution at time $t$,
\begin{eqnarray}
\| \Psi_0(t) - \Psi_0^{\rm exact}(t) \| \le \int_0^t \varepsilon_{\cal M}[\Psi_0(\tau)] d\tau  
\end{eqnarray}
and therefore can be used to minimize the error accumulated in time. Furthermore,
$\varepsilon_{\cal M}^2$ can be understood as a measure of energy fluctuations
accommodated by the variational manifold.\cite{Martinazzo2020,Martinazzo2019}

In fact, the LITE is much more than a numerical tool: it is a {\em quantum
distance} that measures how well a dynamical approximation performs in the
short run. To see this, consider the overlap between the variationally evolved
state and the exact state after an infinitesimal time $dt$,  $S(dt) =
\langle \Psi_0(dt) \vert \Psi_0^{\rm exact}(dt)\rangle$.
This overlap can be connected to a {\em gauge invariant
distance} {$D$, i.e., the so-called Fubini-Study (FS)
distance \cite{Provost1980,Anandan1991,Pati95}}
\begin{equation}
D^{2}(\Psi_0(dt),\Psi_0^{\rm exact}(dt))=2 \left(1- |S(dt)| \right)
\label{eq:FS_overlap}
\end{equation}
which reduces to zero if the variational solution is exact. Within the
geometric interpretation of quantum mechanics,\cite{Anandan1991} the
FS metric is the {natural distance in the {\em
projective} Hilbert space $\mathcal{P}(\mathcal{H})$} whose elements are {\em
physical states} that encompass wavefunctions $e^{i\phi}\Psi$ that differ by
an arbitrary phase factor (in practice, the space of the density operators for pure states).

We will now show, as an extension to Ref.\ [\onlinecite{Martinazzo2020}], that
the {differential} FS distance $d D/dt$, {with
$D$ defined in} Eq.\ (\ref{eq:FS_overlap}), is identical to the LITE of
Eq.\ (\ref{eq:LITE2}). To this end, we expand the overlap up to second order
in $dt$,
\begin{eqnarray}
S(dt) &
 = &
 1+\braket{\dot{\Psi}_0|\Psi_0}dt-\frac{i}{\hbar}\braket{\Psi_0|H|\Psi_0}dt+\nonumber\\
\mbox{} & &
 +\braket{\ddot{\Psi}_0|\Psi_0}\frac{dt^{2}}{2}-\frac{i}{\hbar}\braket{\dot{\Psi}_0|H|\Psi_0}dt^{2}
\nonumber\\
\mbox{} & &
 -\frac{1}{2\hbar^{2}}\braket{\Psi_0|H^{2}|\Psi_0}dt^{2}+\mathcal{O}(dt^{3})
\label{eq:expansion}
\end{eqnarray}
Here, the first order contribution vanishes due to the stationarity condition
with respect to dilations,
$\braket{\dot{\Psi}_0|\Psi_0}=i\hbar^{-1}
\braket{\Psi_0|H|\Psi_0}$ (Eq.\ (4) of Ref.\ [\onlinecite{Martinazzo2020}]). The
same equation shows that
\[
\Re\braket{\ddot{\Psi}_0|\Psi_0}=\Re\left(\frac{d}{dt}\braket{\dot{\Psi}_0|\Psi_0}-||\dot{\Psi}_0||^{2}\right)=-||\dot{\Psi}_0||^{2}
\]
On the other hand, we also have (Eq.\ (6) of Ref.\ [\onlinecite{Martinazzo2020}])
\[
\hbar||\dot{\Psi}_0||^{2}=\Im\braket{\dot{\Psi}_0|H|\Psi_0}
\]
All taken together, Eq.\ (\ref{eq:expansion}) reduces to
\begin{eqnarray}
S(dt)\nonumber & = &
1+\frac{1}{2}\left(||\dot{\Psi}_0||^{2}-\frac{1}{\hbar^{2}}\braket{\Psi_0|H^{2}|\Psi_0}\right)dt^{2}
\nonumber \\
\mbox{} & & + \mathcal{O}(dt^{3})
\label{eq:expansion2}
\end{eqnarray}
Inserting Eq.\ (\ref{eq:expansion2}) into Eq.\ (\ref{eq:FS_overlap}), we find 
up to second order in $dt$, 
\begin{eqnarray}
D^{2}({\Psi}_0({dt}),{\Psi}_0^{\text{exact}}(dt))&=&
\left(\frac{1}{\hbar^{2}}\braket{\Psi_0|H^{2}|\Psi_0}-||\dot{\Psi}_0||^{2}\right)dt^{2}
\nonumber \\
\mbox{} & = & \varepsilon_{\cal M}^2[\Psi_0] dt^{2} \label{eq:equivalence}
\end{eqnarray}
where the term in brackets has been identified as the squared LITE of
Eq.\ (\ref{eq:LITE2}).
Hence, we obtain $\varepsilon_{\cal M} = dD/dt$.

The above derivation also explains the connection with the energy fluctuations
{$\Delta E_0^{2}=\braket{\Psi_0|(H-\braket{H}_0)^2|\Psi_0}$} mentioned above, which is
made evident when working in the standard \emph{gauge},\cite{Martinazzo2020}
where $\Psi_0$ evolves according to the zero-averaged Hamiltonian,
i.e. $\braket{\Psi_0|\dot{\Psi}_0^{+}}=0$,  where the superscript $^+$
denotes the chosen {\emph{gauge}}. This leads to\cite{Martinazzo2020}
\begin{eqnarray} 
\varepsilon_{\cal M}^2[\Psi_0]
=\frac{\Delta
E_0^{2}}{\hbar^{2}}-||\dot{\Psi}_0^{+}||^{2}
\label{eq:D2fluctuation}
\end{eqnarray}
where the energy fluctuation term $\Delta
E_0/\hbar$ is known to drive the {exact} evolution from $t=0$ to $t=dt$
\cite{Anandan1990}
\[
D^{2}({\Psi_0}(0),{\Psi}_0^{\text{exact}}(dt))=\frac{\Delta E_0^{2}}{\hbar^{2}}dt^{2}+\mathcal{O}(dt^{3})
\]
As can be inferred from Eq.\ (\ref{eq:D2fluctuation}), the
variational path becomes exact if the squared variational time derivative
$||\dot{\Psi}_0^{+}||^{2}$ matches the energy fluctuation term $\Delta
E_0^2/\hbar^2$.
In the language of geometric quantum mechanics,\cite{Anandan1990,Anandan1991} both terms
correspond to {\em magnitudes of velocities} in ${\cal P} ({\cal
H})$,
and it is noteworthy that matching these {\em
magnitudes} is sufficient to make the variational solution exact.

To return to the spawning criterion of
Eq.\ (\ref{eq:LITEerror}), an expansion of the variational basis in accordance
with this criterion optimally reduces the mismatch between
$||\dot{\Psi}_0^{+}||^{2}$ and $\Delta E_0^2/\hbar^2$ in
Eq.\ (\ref{eq:D2fluctuation}), leading to {the largest possible reduction of the LITE
upon extension of the variational manifold}.
In the application shown in Ref.\ [\onlinecite{Martinazzo2020}], the LITE was
monitored continuously during the propagation, and its value was used to
decide \emph{when} and \emph{how} to add or remove basis functions.
As emphasized in
Ref.\ [\onlinecite{Martinazzo2020}] and underscored by the above analysis, the
LITE provides a ``natural'' spawning criterion which is firmly rooted in the
variational framework for the approximate solution to the time-dependent
Schr\"odinger equation. 
We anticipate that it will
play a crucial role in future {\em on-the-fly} variational propagation
schemes, bridging between wavefunction methods and local, Gaussian wavepacket
type basis sets.

\newpage
\appendix

\section{Optimal Spawning in MCTDH}
\label{sec:app1}

For completeness, we show here that the theoretical results
of Ref.\ [\onlinecite{MendiveTapia2020}] follow from the general approach
developed in Ref.\ [\onlinecite{Martinazzo2020}], and how they connect to the
spawning (``rate'') operator $\Gamma$ introduced in the
latter reference. To this end, we first notice that when the spaces tangent
to the variational manifold are complex-linear we can write 
the difference between the variational and exact time derivatives as
\begin{eqnarray}
\ket{\Delta\dot{\Psi}_0}& = &\ket{\dot{\Psi}_0}-\ket{\dot{\Psi}_{0}^{\rm
exact}}\nonumber\\
\mbox{}&=&\frac{1}{i\hbar}\left(\mathcal{P}H\ket{\Psi_0}-H\ket{\Psi_0}\right)
\nonumber\\
\mbox{}&=&-\frac{1}{i\hbar}\mathcal{Q}H\ket{\Psi_0}
\label{eq:dpsidot}
\end{eqnarray} 
where $\mathcal{P}$ is the tangent space projector and $\mathcal{Q}$ is its complement,
$\mathcal{Q}=\mathcal{I}-\mathcal{P}$. Furthermore, in MCTDH theory --- when
spawning involves only the $\kappa^{\textrm{th}}$ degree of freedom --- we
have for the difference between the variational time
derivative with/without spawning
\begin{eqnarray}
\ket{\delta_{s}\dot{\Psi}}\equiv \ket{\dot{\Psi}_{s}}-\ket{\dot{\Psi}_0} = -\Omega_{\kappa}\ket{\Delta\dot{\Psi}_0}=\frac{1}{i\hbar}\Omega_{\kappa}\mathcal{Q}H\ket{\Psi_0}
\nonumber \\\label{eq:dsdpsi}
\end{eqnarray}
where $\Omega_{\kappa}$ is a \emph{projector}
\[
\Omega_{\kappa}=\sum_{\alpha}\sum_{J}\ket{\eta_{\alpha}^{(\kappa)}\Phi_{J}^{(\kappa)}}\bra{\eta_{\alpha}^{(\kappa)}\Phi_{J}^{(\kappa)}}
\]
In this expression, the $\ket{\eta_{\alpha}^{(\kappa)}}$'s
are the (orthonormal) single particle functions (spf's) that
are to be added in the spawning process, to be taken from
the orthogonal complement of the space spanned by the spf's of the 
$\kappa^{\textrm{th}}$ degree of freedom. Furthermore, $\ket{\Phi_J^{(\kappa)}}$ is a
configuration exhibiting a ``hole'' at the
$\kappa^{\textrm{th}}$ position and $J$ is a multi-index running over the
occupied spf's of all degrees of freedom but the
$\kappa^{\textrm{th}}$ (see, e.g., Eq.\ (22) in
Ref.\ [\onlinecite{MendiveTapia2020}]).

Upon combining Eq.\ (\ref{eq:dpsidot}) and Eq.\ (\ref{eq:dsdpsi}),
we obtain at once
\begin{eqnarray}
\braket{\delta_{s}\dot{\Psi}|\Delta\dot{\Psi}_0}=-\frac{1}{\hbar^{2}}\braket{\Psi_0|H\mathcal{Q}\,\Omega_{\kappa}\mathcal{Q}H|\Psi_0}
=-||\delta_{s}\dot{\Psi}||^{2}\nonumber\\
\label{dDelta}
\end{eqnarray}
This corresponds to the result obtained in Ref.\ [\onlinecite{MendiveTapia2020}] for the error reduction due to spawning 
\begin{eqnarray}
\Delta\varepsilon_{s}^{2}&=&||\Delta\dot{\Psi}_0||^{2}-||\Delta\dot{\Psi}_0+\delta_{s}\dot{\Psi}||^{2}\nonumber\\
&=&||\delta_{s}\dot{\Psi}||^{2}=\frac{1}{\hbar^{2}}\braket{\Psi_0|H\mathcal{Q}\,\Omega_{\kappa}\mathcal{Q}H|\Psi_0}\label{eq:errorMM}
\end{eqnarray}

On the other hand, from the general theory developed in
Ref.\ [\onlinecite{Martinazzo2020}] (see Eq. (\ref{eq:LITEerror}) of this Comment and, more specifically, Eq. (12) of Ref.\ [\onlinecite{Martinazzo2020}] that applies to the present context), 
we find
\begin{eqnarray}
\Delta\varepsilon_{s}^{2}=\frac{1}{i\hbar}\braket{\delta_s\dot{\Psi}|H|\Psi_0}=\frac{1}{\hbar^{2}}\braket{\Psi_0|H\mathcal{Q}\,\Omega_{\kappa}H|\Psi_0}\label{eq:errorMB}
\end{eqnarray}
which is equivalent to Eq. (\ref{eq:errorMM}) provided
\[
\braket{\Psi_0|H\mathcal{Q}\,\Omega_{\kappa}\mathcal{P}H|\Psi_0}=0
\]
i.e., if and only if
\begin{equation}
\braket{\delta_{s}\dot{\Psi}|\dot{\Psi}_0}=0\label{eq:condition}
\end{equation}
since $\mathcal{P}H\ket{\Psi_0}=i\hbar\ket{\dot{\Psi}_0}$ is the variational
equation of motion. 
Now, the condition Eq. (\ref{eq:condition}) turns out to be 
a consequence of the
Dirac-Frenkel variational condition $\braket{\delta\Psi_0|\Delta\dot{\Psi}_0}=0$.
Indeed, since the latter implies $\braket{\dot{\Psi}_0|\Delta\dot{\Psi}_0}=0$ in
both the original and the extended manifolds, $\mathcal{M}$
and $\mathcal{M}_{s}$, we have
\begin{align*}
0 &
 =\braket{\dot{\Psi}_{s}|\Delta\dot{\Psi}_{s}}=\braket{\dot{\Psi}_{s}|\dot{\Psi}_{s}-\dot{\Psi}_{0}^{\rm
 exact}}\\
 &
 =\braket{\dot{\Psi}_0+\delta_{s}\dot{\Psi}|\dot{\Psi}_0+\delta_{s}\dot{\Psi}-\dot{\Psi}_{0}^{\rm
 exact}}\\
 & =\braket{\dot{\Psi}_0+\delta_{s}\dot{\Psi}|\Delta\dot{\Psi}_0+\delta_{s}\dot{\Psi}}\\
 & =\braket{\delta_{s}\dot{\Psi}|\Delta\dot{\Psi}_0}+\braket{\dot{\Psi}_0|\delta_{s}\dot{\Psi}}+||\delta_{s}\dot{\Psi}||^{2}\\
 & \equiv\braket{\dot{\Psi}_0|\delta_{s}\dot{\Psi}}
\end{align*}
where in the last step we have used Eq. (\ref{dDelta}). 
Hence, Eq. (\ref{eq:errorMM}) and Eq. (\ref{eq:errorMB}) are identical.

In Ref.\ [\onlinecite{MendiveTapia2020}], the error
reduction is given in the form
\[
\Delta\varepsilon_{s}^{2}=\frac{1}{\hbar^{2}}\braket{\Psi_0|H\mathcal{Q}\,\Omega_{\kappa}\mathcal{Q}H|\Psi_0}=\frac{1}{\hbar^{2}}\sum_{\alpha}\braket{\eta_{\alpha}^{(\kappa)}|\Delta^{(\kappa)}|\eta_{\alpha}^{(\kappa)}}
\]
thereby introducing the single-particle operator
\[
\Delta^{(\kappa)}=\sum_J \braket{\Phi_{J}^{(\kappa)}|\mathcal{Q}H|\Psi_0}\braket{\Psi_0|H\mathcal{Q}|\Phi_{J}^{(\kappa)}}
\]
that involves the \emph{many-body} projector $\mathcal{Q}$. The equivalence
between Eq. (\ref{eq:errorMM}) and Eq. (\ref{eq:errorMB}) shows,
quite remarkably, 
that one of the two $\mathcal{Q}$ projectors is irrelevant for the error
reduction and can be safely omitted.

In Ref.\ [\onlinecite{Martinazzo2020}], we further removed
the other $\mathcal{Q}$ projector by requiring additional, simple
constraints on the sought-for spf's, \emph{i.e.}, forcing
them to be orthogonal to both the occupied space \emph{and} its
time-derivative. Under these constraints,
\begin{eqnarray}
\Delta\varepsilon_{s}^{2}&\equiv&\frac{1}{\hbar^{2}}\braket{\Psi_0|H\,\Omega_{\kappa}H|\Psi_0}\nonumber\\
\mbox{}&=&\frac{1}{\hbar^{2}}\sum_{\alpha}\braket{\eta_{\alpha}^{(\kappa)}|\Gamma^{(\kappa)}|\eta_{\alpha}^{(\kappa)}}
\label{eq:desq}
\end{eqnarray}
where 
\begin{eqnarray}
\Gamma^{(\kappa)}=\sum_J
\braket{\Phi_{J}^{(\kappa)}|H|\Psi_0}\braket{\Psi_0|H|\Phi_{J}^{(\kappa)}}
\label{eq:gamma}
\end{eqnarray}
is a much simpler (and therefore computationally cheap)
 generalized ``rate'' operator, i.e., an effective {\em
spawning operator}. Notice that the equality  of
Eq.\ (\ref{eq:desq}) holds \emph{strictly} under the above conditions,
i.e., the
approximation lies in a restricted functional variation. As a result, the
approximation further provides lower bounds to the maximum error reduction
that can be achieved upon spawning
\[
\gamma_{\alpha}^{(\kappa)}\leq\delta_{\alpha}^{(\kappa)}
\]
as follows from the Ritz (Courant-Fischer) theorem when the eigenvalues of $\Gamma^{(\kappa)}$ and $\Delta^{(\kappa)}$
($\gamma_{\alpha}^{(\kappa)}$ and $\delta_{\alpha}^{(\kappa)}$,
respectively) are sorted in decreasing order of magnitude.


%

\end{document}